# PHONON RESONANT CAVITIES AS A PROMISING BUILDING BLOCKS OF HAND-MADE NANOCRYSTALLINE MATTER

V.G.Andreev[1], L.V.Kravchuk[1], S.G.Lebedev[1], A.V.Samokhin[2], and N.V.Alekseev[2]
[1]Institute for Nuclear Research RAS, 117312, Moscow, Russia
[2]Baikov Institute of Metallurgy and Material Science RAS, 119991, Moscow, Russia

**Abstract**

*The nanocrystallite have the finite number of the oscillation modes. Their number increases proportionally to a cube of the characteristic size. Thus the oscillation spectrum of nanocrystal becomes discrete, and the separate modes of oscillations does not interact with each other, that considerably strengthens all phonon modulated processes in a crystal. Covering of such a nanocrystallite with the shielding surface of a material with the higher nuclear weight will allow to create the phonon resonant cavity whose oscillation modes will represent the standing waves and, will be amplified by the resonant manner. The composites made of phonon resonant cavities will allow to produce a perspective functional material for the electronics with adjustable structure and properties.*

The only mode of high-frequency phonon oscillations that can be obtained in metallic devices made with the use of conventional manufactoring processes is that of travelling wave. Under these conditions any perturbation at a point within the crystal lattice will travel with dissipation in all directions from that point and never come back. In a travelling wave mode phonon oscillation fields are random. Hence, the interaction between conduction electrons and the oscillations of the crystal lattice of an electrical conductor has a viscous nature. We propose to replace the travelling wave mode with the standing resonant wave mode for frequencies from 0.1 to 1 *THz* and higher.

The properties of polycrystalline materials substantially depend on their crystallite size [1]. With decreasing of the crystallite size the number of constituent atoms decreases proportionally to a cube of the linear size. With decreasing of atoms number in the crystallite the number of oscillation modes of a crystal lattice also decreases. Nanocrystallite with the size of 2-3 nanometers consist of $N \sim 10^2$ atoms. The full number of oscillation modes in the lattice of such nanocrystallite is *3N*. At such small number the oscillation modes are isolated each from other and do not interact with each other. In that case the oscillation spectrum of separate nanocrystallite is completely harmonic. The composite material consisting from such nonocrystallites, will possess the unique properties. For example, such composite will not be subjected to thermal expansion (due to absence of the unharmonicity), the plasticity and the strength of such alloy will sharply increase as high as 8-10 times, the melting point decreases as compared with the microcrystalline composite. But much more interesting changes, in our opinion, will be possible to expect in the electrodynamics of such nanocrystalline composites. From this point of view the composites made of nanocrystallites which external surface is covered with the thin (in some atomic layer) shielding surface made of element with higher atomic number may be of interest. This gives rise to reflection of a sound wave from a shielding surface and formation of resonant standing wave inside the nanocrystallite. Such "packed" nanocrystallite has been called by the phonon resonant cavity [2]. The upper cutoff frequency of a phonon cavity is determined by the mass of atoms of the nanocrystalline lattice, while the lower cutoff frequency depends upon the screen atoms mass. The screen is transparent in the band between the upper frequency of nanocluster's atoms and the lower frequency of screen's atoms. Hence no standing wave can be excited in this range but only the travelling wave. Standing wave can only be excited in the range below the lower cutoff frequency. Till now the dynamics of oscillations of a lattice has been connected only with the materials based on the travelling wave propagation. As it is well known the resonant oscillations are extremely amplified on a background of other frequencies, therefore all phonon modulated processes will be strengthened.

However the covering of nanoclusters with the nanosceens is rather not trivial problem [3-4]. Problems arise due to both the high mobility of nanoclusters - they intensively "stick together" during the deposition from a vapor phase, and also the violent interaction with the atmospheric oxygen. The oxidation of nanocrystallites in an atmosphere can proceeds by the explosive manner due to their extremely developed surface, therefore during deposition from a vapor the measures should be taken on fast and effective quenching of nanoclusters. The subsequent processes of a nanoclusters covering with the nanoscreens and the process of sintering of nanocomposite from a powder of phonon resonant cavities should occur without access of the oxygen. As a result of all listed actions we can obtain the first rather man-made nanocomposite with the controllable properties - the size of grains and the thickness of layers. So far people faced only with the nanocomposites presented by the nature which properties frequently are casual and are not adjustable. It is necessary to expect, that nanocomposites of phonon resonant cavities (*NPRC*) will possess an interesting electrodynamic properties. As phonon modes in the nanocrystallites are independent, they cannot interfere and do not attenuate each other, and, therefore it is necessary to expect the strengthening of such phonon attributed processes as the superconductivity. Thus, it is hoped to raise essentially the temperatures of superconducting transition in *NPRC*. Till now in the world practice there are no of such kind of works and the influence of dimensional effects on superconductivity is studied only on an example of thin films and nanowires where the effects expected considerably concede as compared with those of *NPRC*.

Other interesting aspect of *NPRC* is the opportunity to simulate the nanostructured josephson media, which is the subject of our works [5-10] where the attention has been payed to anomalous electromagnetics in the nanosized josephson-like structure of granular carbon films. It is believed, the structure of this films consists of the superconductive graphite granules embedded in a matrix of amorphous carbon with the high electroresistance. Thus the sizes of the granules and intergranular layers have a wide scatter that leads to formation of finite superconducting clusters according to percollation mechanism of conductivity in a composite. Production of controllable *NPRC* will allow to minimize the spread in the sizes of their granules and layers and to create the nanocomposite superconductor with the josephson link. And there is the possibility to vary the sizes of granules and thickness of layers by the controllable manner and consequently this allow to adjust the value of josephson link between the granules. This gives rise to large variety of possible applications as the logic and memory elements of quantum computers, detectors [11] and generators of the microwave radiation, which can be considered as the essentially new products and prospective objects for intellectual property. Optimization of the structure of the granular *NPRC* with the purpose of strengthening of the effects specified above will allow to develop the new extremely prospective functional material for creation of the devices of josephson-like electronics. The activity on creation of *NPRC* are lie on a border line of modern nanoelectronics around the world.

**ACKNOWLEDGEMENTS**

The authors would like to thank to Academician of RAS V.A.Matveev for support of this work and we are gratefully acknowledge the financial support from the Russian Foundation of Basic Research through Grant No. 07-08-13516-офи-ц.